\documentclass[article,singlecolumn,superscriptaddress,nopacs,floatfix,amssymb,amsmath]{revtex4}
\usepackage{epsfig}
\usepackage{graphics}
\usepackage{array}
\usepackage{multirow}

\newcommand{\beqn}{\begin{eqnarray}}
\newcommand{\eeqn}{\end{eqnarray}}

\begin{document}

\title{Towards Quantitative Classification of  Folded Proteins \\ in Terms of  Elementary Functions}

\author{Shuangwei Hu}
\email{Shuangwei.Hu@lmpt.univ-tours.fr}
\affiliation{
Laboratoire de Mathematiques et Physique Theorique
CNRS UMR 6083, F\'ed\'eration Denis Poisson, Universit\'e de Tours,
Parc de Grandmont, F37200, Tours, France}
\affiliation{Department of Physics and Astronomy, Uppsala University,
P.O. Box 803, S-75108, Uppsala, Sweden}
\author{Andrei Krokhotin}
\email{Andrei.Krokhotine@cern.ch}
\affiliation{Department of Physics and Astronomy, Uppsala University,
P.O. Box 803, S-75108, Uppsala, Sweden}
\author{Antti J. Niemi}
\email{Antti.Niemi@physics.uu.se}
\affiliation{
Laboratoire de Mathematiques et Physique Theorique
CNRS UMR 6083, F\'ed\'eration Denis Poisson, Universit\'e de Tours,
Parc de Grandmont, F37200, Tours, France}
\affiliation{Department of Physics and Astronomy, Uppsala University,
P.O. Box 803, S-75108, Uppsala, Sweden}
\author{Xubiao Peng}
\email{xubiaopeng@gmail.com}
\affiliation{Department of Physics and Astronomy, Uppsala University,
P.O. Box 803, S-75108, Uppsala, Sweden}

\begin{abstract}
A comparative  classification scheme provides a good basis for several approaches 
to understand proteins, including prediction of relations between their structure and biological  function.  
But  it remains a challenge to combine a classification scheme that describes a protein 
starting from its well organized secondary structures and often involves direct human 
involvement, with an atomary level Physics based approach where a protein is fundamentally nothing more
than an ensemble of mutually interacting carbon, hydrogen, oxygen and nitrogen atoms.  
In order to  bridge these two complementary approaches to proteins, conceptually novel 
tools need to be introduced.  Here we explain how  the geometrical shape of  
entire folded proteins  can be described analytically in terms of a 
single explicit elementary function that is familiar from nonlinear physical systems where it is known as 
the kink-soliton. Our approach enables the conversion of  hierarchical structural information into a 
quantitative form that allows for a folded protein to be characterized in terms of a small number of 
global parameters that are in principle computable from atomary level considerations. As an example 
we describe in detail how the native fold of the myoglobin 1M6C emerges 
from  a combination of kink-solitons with a very high atomary level  accuracy. We also 
verify that our approach describes longer loops and loops connecting $\alpha$-helices with $\beta$-strands,
with same overall accuracy.
\end{abstract}

\maketitle

\section{Introduction}
Comparative protein classification schemes such as CATH \cite{cath} and SCOP \cite{scop} 
are among the most valuable and widely employed
tools in bioinformatics based approaches to protein structure. These schemes 
classify  folded proteins in terms of their geometric shape, 
starting from  prevalent secondary structures such as $\alpha$-helices and $\beta$-strands. 
But at the moment the final stages of the classification usually involve manual curation, and 
consequently these schemes are best  suited for qualitative analysis  of folded proteins.

The goal of the present article is to develop  novel tools that we propose can eventually provide a firm
quantitative   basis for the existing protein classification schemes.
Ultimately we hope to close gaps between bioinformatics 
based protein structure classification and physics based 
atomary level approaches to protein folding, to comprehensively address  wide range of
 issues such as protein structure prediction and relations between
shape, function and dynamics. In this way we hope to open doors to new  ways 
to perform evolutionary, energetic and modelling studies.

Our approach is based on the recent observation \cite{cherno}, \cite{nora} that the geometric shape of helix-loop-helix
motifs  can be captured by a single elementary function 
that is familiar from the physics of nonlinear systems
where it describes the kink-soliton. This function involves only a relatively 
small set of global parameters but still characterizes an entire super-secondary structure 
involving two ($\alpha$-)helices and/or  ($\beta$-)strands in addition of the loop that connects them. 
In \cite{cherno} only individual supersecondary structures in relatively simple proteins and with quite short loops were considered. 
The approach proposed there did not work very well for entire protein chains, involving several helices and loops, it
was essentially limited to a relatively short single loop with adjoining helices. The purpose of the present article is to show that the method 
can be developed to describe  an {\it entire} protein and not just its helix-loop-helix segments. The protein
can also be quite complex, it can involve several loops, both short and long and 
including those that connect $\alpha$ helices with $\beta$ strands. 
Furthermore, the  original Ansatz can be even simplified  without affecting  its accuracy. 
Remarkably we observe no loss of accuracy even when 
the length and complexity of the protein chain increases. Indeed, there does not appear to be any limitations whatsoever 
that have to be imposed on the complexity of the protein, for our approach to remain practical.

Our motivation derives from an investigation  of nonlinearities that are generic in the force
fields employed in classical molecular dynamics,  a  technique  that is widely used in various 
theoretical studies of the structure,  dynamics and thermodynamical properties  of  proteins,
and  in determining their folding patterns in x-ray crystallography and NMR experiments  \cite{md}.  
A classical molecular dynamics approach like  AMBER  \cite{amber}  and GROMACS \cite{gromacs} describes the evolution 
of a folding protein in terms of Newton's law that determines the time dependence of  the  atomary
spatial  coordinates 
$\mathbf X(t) = \{ \mathbf x_i(t)\}  $ 
\begin{equation}
m_i  {\bf \ddot x}_i (t) \ = \ - \nabla_i U(\mathbf X) 
\label{new}
\end{equation}
Here $i=1,...,N$ catalogue  the individual atoms both in the protein molecule and its environment, 
and $U(\mathbf X)$  is an empirically constructed potential energy that governs the relevant mutual 
interactions between all atoms involved.

Generically the potential energy  is written as  the sum of two terms \cite{amber}
\begin{equation}
U(\mathbf X) = \sum U_{\text{covalent}}(\mathbf X) + \sum U_{\text{rest}}(\mathbf X)
\label{Utot}
\end{equation}
The first term 
describes the covalent  two-, three-, and four-body interactions between all covalently bonded atoms.
The second  term  
describes  the non-covalent interactions between all atoms. 
For example, in the widely used harmonic approximation the two-body contribution to potential energy
that describes the vibrational 
motion of all pairs of covalently bonded atoms acquires the familiar form 
\begin{equation}
U^{(2)}_{\mathrm{bond}} = \sum_{\mathrm{bonds}} k_{ij} ( |\mathbf x_i - \mathbf x_j| - r_{0ij})^2
\label{U2}
\end{equation}
where $r_{0ij}$ are the equilibrium distances between the pairs of covalently bonded atoms $i$ and $j$, and $k_{ij}$ are 
the ensuing spring constants.  

But there are also nonlinear corrections to the potential energy such as (\ref{U2}), albeit in 
practice they may be difficult to account for in a systematic manner. The study of these nonlinearities forms a basis of
the present work. 

We start with a {\it Gedanken} experiment where we scrutinize a highly simplified version 
of an improvement to the  harmonic approximation (\ref{U2}),  with only a single (relative) coordinate on a line $x$  so that Newton's equation 
is mere
\[
m \ddot x =  - \frac{dV}{dx} 
\]
where the potential has the form
\[
V(x) = \frac{1}{2} k(x) \cdot  (x-a)^2 \approx \frac{1}{4} \hskip 0.3mm \kappa \, (  x+b)^2 \cdot  (x-a)^2
\]
That is we account for nonlinear deviations from the harmonic approximation 
by promoting the spring constant to a $x$-dependent quantity. The equilibrium position  $x=a$ of the harmonic 
approximation  is recovered when $|x| \approx |a| << |b|$, 
\[
V(x) \ \approx \ \frac{1}{4} \kappa \hskip 0.5mm b^2 (x-a)^2  \cdot \left( 1 + \mathcal O(\frac{x}{b}) \right)
\]
but here we retain the full potential.  We introduce
 \[
 c =  -\frac{1}{2} (b+a)
 \]
 and define 
 \[
 y  = x -  \frac{1}{2}(a-b)
 \]
 to arrive at the familiar "$\lambda \phi^4$" (kink) equation of motion
 \[
 \ddot y  =  - \frac{\kappa}{m} { y} ({y}^2 - c^2)
 \]
 with the explicit dark soliton solution
\[
 { y }(t) \ = \  c \cdot \tanh [\, c \, \sqrt \frac{ \kappa}{2m}  (t-t_0) ]  
 \]
 \[
  \Rightarrow \  x(t) \ = \  y(t) + \frac{1}{2} (a-b) 
  \]
   \begin{equation}
 = \ - \, \frac{ b \cdot  e^{c \, \sqrt \frac{ \kappa}{2m}  (t-t_0) } \ -  \  a  \cdot
 e^{- c \, \sqrt \frac{ \kappa}{2m}  (t-t_0) } } {  \cosh [ c \, \sqrt \frac{ \kappa}{2m}  (t-t_0) ] }
 \label{dark}
 \end{equation}
This is the hallmark dark soliton (kink) configuration that
interpolates between the two uniform ground states at $x = a $ and $x=-b$ when $t \to \pm \infty$. 
The parameters $a$, $b$, $t_0$ and the combination $c \, \sqrt \frac{ \kappa}{2m}$ are the canonical ones that characterize the asymptotic values
of $x(t)$ {\it i.e.} minima of the potential,  and the size and location of the soliton.
It is also noteworthy that for finite $t$ the soliton (\ref{dark}) describes  a configuration with an energy
above  the uniform ground state $x\equiv a$  (or $x \equiv b$) but  that nevertheless can not decay into  $x  \equiv a$ (or $x \equiv b$)
through any kind of continuous finite energy 
transformation: A soliton configuration such as (\ref{dark}) 
can not be obtained from any approach that only accounts for perturbations that
describe small localized  fluctuations around the uniform background ground state.  

We argue that our example is not just an academic exercise but can be developed into a systematic 
tool to quantitatively  characterize the geometrical shape of super-secondary structures such as helix-loop-helix motifs. In fact,
we propose that
the {\it very same function} (\ref{dark}) with $t$ a length  parameter  
that measures distance along a static protein backbone, together with its asymmetric generalization of the form
\begin{equation}
\tilde x(t) =  \frac{ 
b \cdot  e^{\, c_1 (t-t_0) } \ -  \  a  \cdot
 e^{- c_2  (t-t_0) } 
 } {   
e^{c_1  (t-t_0) } \ +  \ 
 e^{- c_2   (t-t_0)}
  }
\label{dark2}
\end{equation}
which becomes handy {\it e.g.} when we consider loops connecting an $\alpha$ helix with a $\beta$ strand,
can describe the geometry of native folds of proteins  in Protein Data Bank (PDB) \cite{pdb}. 
Besides  the four canonical soliton parameters that  we have specified,  we need  to introduce only two
additional  independent global parameters to characterize a 
given super-secondary structure such as a helix-loop-helix  motif and even an entire folded protein, 
with an  atomary level accuracy  that matches the resolution in experimental data.

As an explicit example we have chosen myoglobin, a widely studied oxygen-binding protein of
both historical and biological interest that has been discussed extensively in
most biochemistry texts. Specifically, we have selected the 
153 amino acid  myoglobin with Protein Data 
Bank code 1M6C whose all-atom structure is known to an all-atom resolution 
of 1.90 \.A in root-mean-square distance (RMSD) from  x-ray diffraction measurements \cite{pdb}.   
We analyze it in detail, to show that  its entire fold can be encoded into the global parameters of the elementary function (\ref{dark}), (\ref{dark2})
with a RMSD accuracy of 1.27 \.A for the central $C_\alpha$ carbons. Moreover, as  the myoglobin only involves
super-secondary structures with $\alpha$ and $3/10$ helices that are connected by  relatively short loops, we also verify that our approach can be 
extended to longer loops, and loops that  connect $\alpha$ helices with $\beta$ strands. For this we  analyze an $\alpha$ helix - loop - $\beta$
strand segment in the HIV-1 reverse transcriptase protein with PDB code 3DLK. The loop is now clearly longer than those in
myoglobin, nevertheless  we find that it  can be described with comparable RMSD accuracy by the
profile (\ref{dark2}).

\section{Myoglobin as multisoliton}

In order to describe the PDB  fold of  a relatively complex protein such as the $153$ amino acid 1M6C in
terms of the {\it single} elementary function (\ref{dark}),
we start by computing the values of its  discrete Frenet curvature $\kappa_i  $ 
and Frenet torsion $\tau_i $ from the PDB data. The relevant equations are as follows (for detailed derivation, see \cite{martin}): 
From PDB  we get the three dimensional coordinates $\mathbf r_i$ 
of the  central $\alpha$-carbons ($i=1,...,N$).  With these we compute the tangent
vector $\mathbf t_i$ and the binormal vector $\mathbf b_i$ using
\begin{equation}
\begin{matrix}
\mathbf t_i = \frac{ \mathbf r_{i+1} - \mathbf r_i}{ | \mathbf r_{i+1} - \mathbf r_i |}  \ \ \ \ \& \ \ \ \ \
\mathbf b_i =
\frac{ \mathbf t_{i-1} \times \mathbf t_i}{ | \mathbf t_{i-1} - \mathbf t_i |}
\end{matrix}
\label{tb}
\end{equation}
and the normal vector is given as
\[
\mathbf n_i = \mathbf b_i \times \mathbf t_i
\] 
These three vectors are subject to  the discrete Frenet equation  
\begin{equation}
\left( 
\begin{matrix} 
{\bf n}  \\  {\bf b }  \\  {\bf t}  
\end{matrix} \right)_{i+1}  =
\exp\{ - \kappa_{i} \cdot T^2 \} \cdot \exp \{ - \tau_i \cdot T^3 \}
\left( \begin{matrix} {\bf n} \\  {\bf b } \\ {\bf t} \end{matrix} \right)_i
\label{frenet}
\end{equation}
Here $T^2$ and $T^3$ are two of the standard adjoint generators of three dimensional rotations, explicitly in terms of the
permutation tensor we have 
\[
(T^i)^{jk} = \epsilon^{ijk}
\]
From (\ref{tb}), (\ref{frenet}) we can  compute $\kappa_i$ and $\tau_i$ as the bond angles and the torsion angles in terms of
the PDB data for $\mathbf r_i$. Alternatively, if we know these angles we can compute the coordinates $\mathbf r_i$ up to global
rotations and translations.
The common convention is to select  the range of these angles so that  $\kappa_i$ is non-negative. In the
continuum limit where (\ref{frenet}) becomes the standard  Frenet equation for a continuous curve,
$\kappa_i \to \kappa(x)$ then corresponds to local curvature which is  by convention defined to be non-negative. 

For 1M6C we take $i$ to take values $i=3,...,149=N$;  We leave out three (four) sites at both end as we need  three sites  
to initiate the computation of the $\kappa_i$ and $\tau_i$ along the polygon, and  the end points are anyway presumed to be subject to
relatively large conformational fluctuations. 
In Figure 1 (top) we display the
$\kappa_i$ and $\tau_i$ along the myoglobin backbone, using the standard differential
geometric convention that $\kappa_i$ is non-negative. 
%
\begin{figure}[h]
\includegraphics[scale=0.55,clip=true]{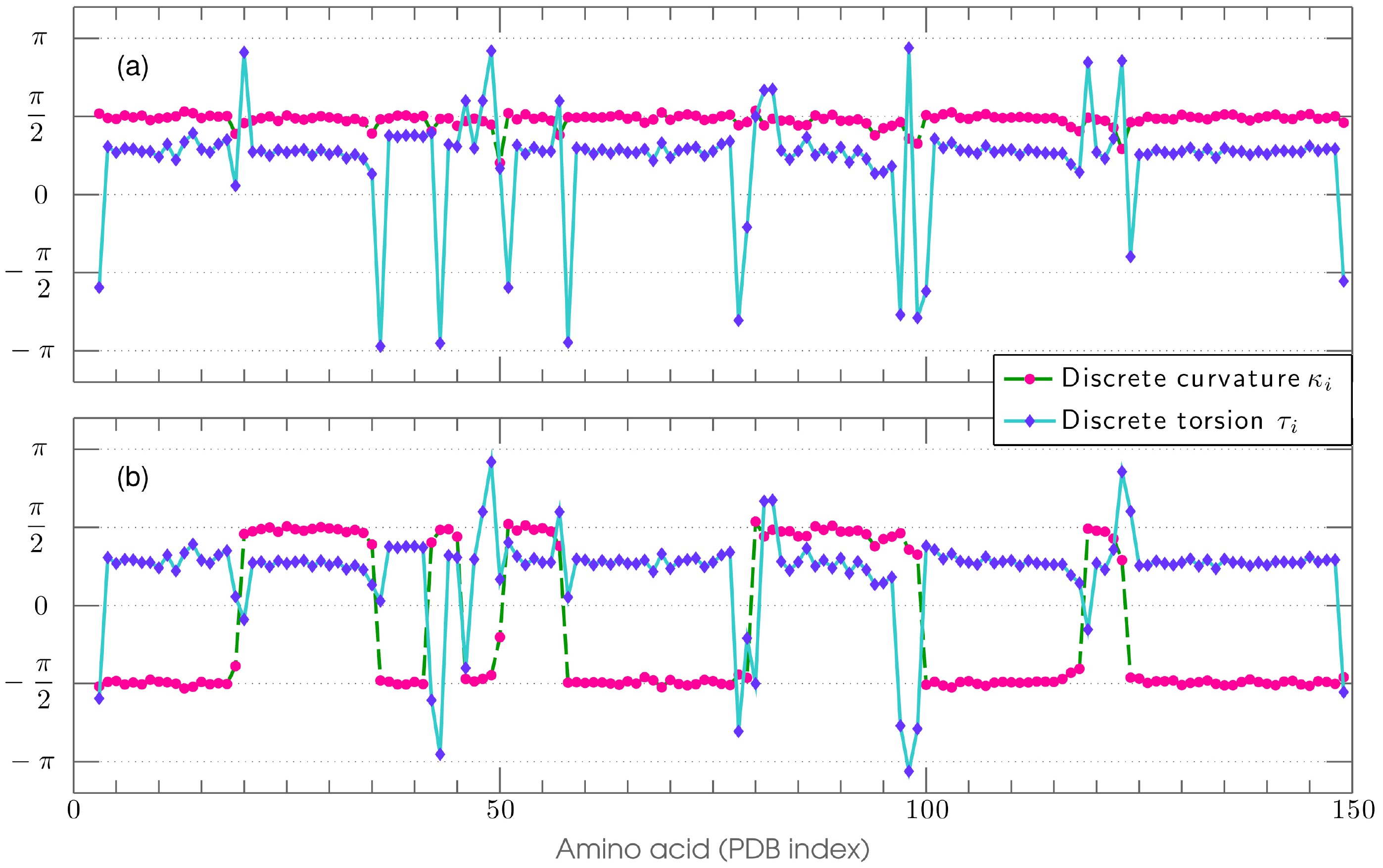}
\caption{The  values of $\kappa_i$ and $\tau_i$ for 1M6C, obtained from PDB. In the top picture we present 
these values using the standard convention that  $\kappa_i$ is non-negative. In the bottom
picture we have resolved the soliton structure using $\mathbb Z_2$ gauge structure of the Frenet equation, 
by allowing $\kappa_i$ to change sign whenever there is an inflection point. 
This identifies the soliton structures (loops) along the backbone. The indexing refers to the 
position of amino acids along the backbone, counting from the $N$-terminus.}\label{fig:01}
\end{figure}
This Figure displays the geometric structure of the 1M6C backbone fold: At the location of the $\alpha$ and $3/10$ helices both $\kappa_i$ and $\tau_i$ have
pretty constant values, as expected for helical geometry. The difference between these two types of helices is visible in the Figure, in (slight) difference
in the corresponding constant  values of $\kappa_i$ and $\tau_i$. At the location of loops, we note small variations in $\kappa_i$
while the values of $\tau_i$ are fluctuating quite wildly. 
In order to identify the locations of the
inflection points that determine the center of the loops {\it i.e.} solitons, we follow  \cite{cherno}  and
subject the data in Figure 1 (top) to local $\mathbb Z_2$ gauge transformations in the loop regions; these transformations
leave the solution of (\ref{frenet}) intact
and thus have no effect on the geometry of the
space polygon.
The result is  shown in Figure 1 (bottom); the two data point sets in the top and bottom of Figure 1 describe
the same space polygon.  But from the bottom Figure 1 we conclude that in terms of $\kappa_i$  we may 
interpret the backbone as a space polygon with eleven helices that are separated by ten inflection points (soliton centers), these are the
points where $\kappa_i$ changes its sign.   Consequently we divide the backbone into ten super-secondary 
structures,  each consisting of a  helix-loop-helix soliton motif.  These motifs  are identified in Table I.

We note that PDB lists 1M6C as an eight-helix protein. But Figure 1 reveals that there is an advantage to interpret it in terms of
a curve with 
ten inflection points, so that for a match with the functional form (\ref{dark}) we need to introduce ten overlapping 
segments. Furthermore, an examination of the PDB data reveals  that  
there are four different types of  loops {\it i.e.} solitons:  Those that connect
two $\alpha$ helices, those that connect an $\alpha$-helix with a $3/10$-helix or {\it vice versa}, 
and finally those that connect two $3/10$-helices. 

In order to describe a motif consisting of a loop together with the  two similar types of helices that it connects, we  use the
Ansatz (\ref{dark}) with the symmetric  ($a=b$)  relation for the two parameters in (\ref{dark}). 
But for motifs where a loop connects two different types of helices ($\alpha$ with $3/10$) 
we allow these parameters to be independent, reflecting the
difference in the helices.
Thus our Ansatz for the entire backbone is  the  modification (\ref{dark2})  of the Ansatz introduced
in \cite{cherno}: For the bond angles we introduce the dark soliton profile
\begin{equation}
\kappa_i \ = \  (-1)^{r+1}  \frac{ 
m_{r 1}  \cdot e^{ c_r ( i-s_r) } - m_{r2} \cdot e^{ - c_r ( i-s_r)}  }
{
2 \, \cosh[  c_r ( i-s_r)] }
\label{An1}
\end{equation}
and we obtain the torsion angles from this soliton profile using the relation
\begin{equation} 
\tau_i \ = \ - \frac{1}{2} \frac{ b_r }{1  + d_r \kappa_i^2}
\label{ti}
\end{equation}
Here $r=1,...,10$ labels the ten helix-loop-helix motifs of 1M6C and $ (c_r, m_{r1}, m_{r2}, s_r)$ are the canonical parameters for a kink-soliton,
and $ (b_r , d_r)$ are additional parameters needed to express $\tau_i$ in terms of $\kappa_i$. 

Note that  in (\ref{ti}) we have  {\it simplified} the Ansatz  of \cite{cherno} for torsion angles. Now there is no contribution from $\kappa_i$ in
the numerator, thus there is one less parameter.  The reason for this simplification is, that (\ref{An1}), (\ref{ti}) is {\it not} an
{\it ad hoc} Ansatz but can be firmly  justified in terms of the equations of motion in an underlying Hamiltonian model which is based
on the Abelian Higgs Model \cite{ulf} .  The additional
term  used  in \cite{cherno}  version of (\ref{ti}) does not have any natural interpretation in terms of the Abelian Higgs Model and thus there
is no geometric reason for including it. Here we confirm
that it can be safely removed, with {\it no} adverse effect in accuracy. In fact, despite the additional increased complexity in protein structure
that we consider, the accuracy reported here is  even  better than that in \cite{cherno}.

We also emphasize that the parameters 
are all {\it global } parameters that are 
specific to a given helix-loop-helix motif and {\it as such} have no direct reference to the amino acids even though they should eventually
become computable from an atomary level set-up. At the level of the Abelian Higgs Model \cite{ulf} each of the parameters has a 
well established interpretation in terms of charge, mass, self-coupling {\it etc.}
Here, these parameters characterize the global attributes such as the location and the size of the 
soliton-loop in terms of $\kappa_i $ and $\tau_i$,  the nature of the adjacent 
helices, and the chirality of the protein.  
Moreover, since all the solitons except  $2$ and $5$ connect similar helices, 
whenever $r \not= 2,5$ we can set $m_{r1} = m_{r2}$
while for solitons number $2$ and $5$ that connect two different kind of helices we choose
$m_{r1} \not= m_{r2}$. 
We also  emphasize that the 
Ansatz involves only the {\it single} function (\ref{dark}), in its discrete form. This means that
for each helix-loop-helix superstructure we only need to determine the five (or six in case the helices are different)  global parameters. 
In our computations we determined these parameters 
using a standard Metropolis algorithm 
in combination with simulated annealing,  to minimize
the RMSD between the polygon described by our Ansatz  and  the $C_\alpha$ backbone of 
the 1M6C protein in PDB.  The actual algorithm is a very simple and straightforward application of standard
Monte Carlo minimization that runs  with PC.

In Table I we display the parameters that yield the smallest 
RMSD value  (RMSD = 1.27 \.A)  that we have obtained when we have subjected the entire  
1M6C backbone to a RMSD minimization.
{
\begin{table}[h]
\caption{The parameters for solitons along the 1M6C $C_\alpha$-backbone,
with indexing starting from the $N$ terminus. }\vskip 0.5cm
\begin{tabular}{|c|c|c|c|c|c|}
\hline 
soliton 
 & 1 & 2  & 3  & 4  & 5  
\tabularnewline 
\hline 
sites &  3-24 & 22-42 & 37-46  & 43-50 & 47-58 
\tabularnewline \hline
type & $ \alpha$-$\alpha$ & $\alpha $-3/10   & 3/10-3/10 & 3/10-3/10  & 3/10-$\alpha $ 
 \tabularnewline  
\hline
$b_r$ & 78.398 & 79.1807 & 68.7412  & 39.727 & 55.9241
\tabularnewline\hline
$c_r$ & 1.5708 & 2.5280 & 2.5290 & 2.5550 & 3.1391
\tabularnewline
\hline
$d_r$     & -0.2905 & -0.1268 & -0.2347 & -0.2464 & -0.2998 
\tabularnewline\hline
$m_{r1}$  & 1.53668 & 1.4979 & 1.56503 &  1.55474 & 1.5668
\tabularnewline\hline
$m_{r2}$ & $- $ & 1.5113 & $-$ & $-$ & 1.5651 
\tabularnewline\hline
$s_r$ & 20.5981 & 36.488 & 43.3982 & 45.657 & 51.733
\tabularnewline\hline
rmsd & 0.83 & 0.49 & 0.15 & 0.56 & 0.40
\tabularnewline\hline
\hline 
soliton 
 & 6  & 7  & 8 & 9  & 10 
\tabularnewline 
\hline 
sites &
52-80 &  \hskip 0.0cm 59-98  \hskip 0.0cm &  \hskip 0.0 cm 81-119  \hskip 0.0cm &  \hskip 0.0cm 102-123  \hskip 0.0cm &  \hskip 0.0cm 120-150  \hskip 0.0cm 
\tabularnewline \hline
type &
 $\alpha$ - $\alpha  
$ & $\alpha$-$\alpha$ & $\alpha$-$\alpha$ & $\alpha$-$\alpha$ & $\alpha$-$\alpha$
 \tabularnewline  
 \hline 
$b_r$ & 73.358 & 92.551 & 48.059 & 114.599 & 93.2733 
\tabularnewline\hline
$c_r$ & 2.1488& 2.1874 & 1.95991 & 2.2796 & 2.5496
\tabularnewline\hline
$d_r$ & -0.3035 & -0.4649 & -0.3688 & -0.1887 & -0.1565 
\tabularnewline\hline
$m_{r1}$ & 1.52541 & 1.52732 & 1.48823 & 1.55946 & 1.54715 
\tabularnewline\hline
$s_r$ & 57.8112 & 80.7367 & 98.2245 & 118.8551 & 124.404
\tabularnewline\hline
\hline
rmsd & 1.12 & 1.46 & 1.62 & 0.60 & 0.37
\tabularnewline\hline
\end{tabular} 
\\ \vskip 0.2cm { 
The solitons have some overlap with their nearest
neighbors, to enable us  to combine them into a single multi-soliton profile. The type  
identifies whether the soliton consists of a loop 
that connects $\alpha$-helices and (or) $3/10$-helices.  }
	\label{tab:para2}
\end{table}

}
\noindent
We also give the lowest RMSD 
values that we have obtained when we have separately 
optimized the parameters for each of the individual soliton. For the solitons $1,2,3,4,5,9$ and $10$ we find 
very low RMSD values, clearly smaller than the
radius ($\sim 0.7$ \.A) of an individual carbon atom. However, the number of sites that appear 
in the solitons $3,4,5$ are also quite small. This is due to the proximity of the
ensuing solitons along the backbone. For solitons number $6,7,8$ the RMSD values are somewhat larger, but the solitons are also longer.
However, even in these cases our RMSD
values are clearly below the overall 1.90 \.A resolution in the underlying PDB data.
In Figure 2 we display the $C_\alpha$ backbone of 1M6C, together with its reconstruction in terms of the Ansatz (\ref{An1}), (\ref{ti}).
%
%
%
\begin{figure}[h]
\includegraphics[scale=0.3,clip=true]{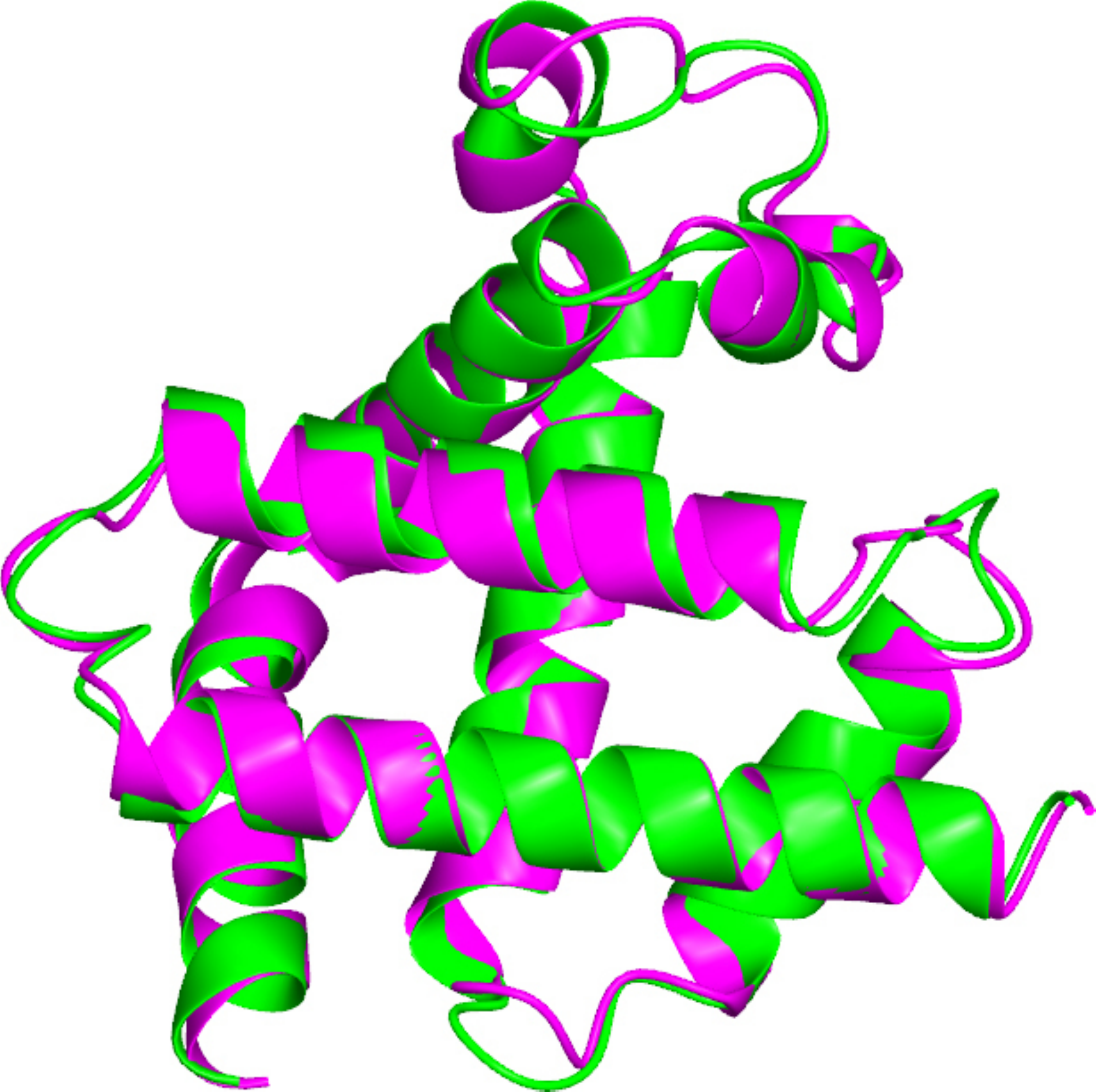}
\caption{The structure of the 1M6C protein (green) together with its reconstruction
in terms of our Ansatz (purple). The RMSD distance between the two configurations is $\approx$ 1.27 \.A.}\label{fig:02}
\end{figure}

We remind that even though our description involves five (six) free parameters for each helix-loop-helix motif,  there is only {\it one single function},
the kink-soliton (\ref{An1}).   These parameters can in principle be determined 
from a first principle  atomary level  approach to protein folding, even though in practice this is not yet possible. For this
we recall \cite{cherno} that as such, (\ref{An1}), (\ref{ti}) is an approximate solution to a definite 
discrete nonlinear equation of  motion in a Hamiltonian system that provides an effective
description of a more fundamental atomary level model. 

\section{long loops}

The previous interpretation and construction of the myoglobin 1M6C backbone clearly demonstrates that the method proposed in \cite{cherno}
can be extended from helix-loop-helix super-secondary structures to entire proteins, even for relatively long proteins and 
with several helix-loop-helix combinations and  both $\alpha$ and $3/10$ helices. However, the question remains whether the quality
of the method becomes adversely affected if the loop length increases, and whether the method also describes loops that connect
an $\alpha$-helix with a $\beta$ strand. We address these issues by considering a protein loop with 12 $C_\alpha$-carbons connecting
an $\alpha$-helix with a $\beta$-strand. More specifically, we consider the sites 398-416 in the HIV-1 reverse transcriptase protein 
with PDB code 3DLK.  In line with the construction of the solitons in the case of myoglobin, we describe the super-secondary structure 
 with the following variant (\ref{dark2}) of the Ansatz (\ref{An1}), (\ref{ti}),
 \begin{equation}
\kappa_i \ = \    \frac{ 
m_{1}  \cdot e^{ c_1 ( i-s_0) } - m_{2} \cdot e^{ - c_2 ( i-s_0)}  }
{
e^{ c_1 ( i-s_0) } +   e^{ - c_2 ( i-s_0)} 
}
\label{An2}
\end{equation}
and we again obtain the torsion angles from this soliton profile using the relation
\begin{equation} 
\tau_i \ = \ - \frac{1}{2} \frac{ b }{1  + d  \kappa_i^2}
\label{ti2}
\end{equation}
The asymmetric choice $(m_1, c_1)$ {\it vs.} $(m_2, c_2)$ reflects the difference between the $\alpha$-helix and
$\beta$-strand, and we now start the indexing by choosing $i=1$ for the site 398. With the choice of parameters in Table II
{
\begin{table}[h]
\caption{The parameters for describing the sites 398-416 along 3DLK. Indexing starts with $i=1$ at site 398.
} \vskip 0.5cm
\begin{tabular}{|c|c|c|c|c|c|c|}
\hline 
$m_1$& $c_1$  & $m_2$  & $c_2$  & $s_0$  &$ b$  & $d $ 
\tabularnewline \hline
57.626008  & 1.836469   & 58.05348 & 1.8462217  & 10.43150  & 6601165.9 & -0.000101 
\tabularnewline\hline
\end{tabular} 
	\label{tab:para3}
\end{table}
}
we find that the Ansatz describes the 3DLK segment with a RMSD  accuracy of 1.13 \.A; Notice that  due to the presence of exponentials,
 for high accuracy it is imperative to include sufficiently many decimal points in the parameters. In Figure 3 we display the original
 3DLK segment, together with its soliton approximation.
 \begin{figure}[h]
\includegraphics[scale=0.3,clip=true]{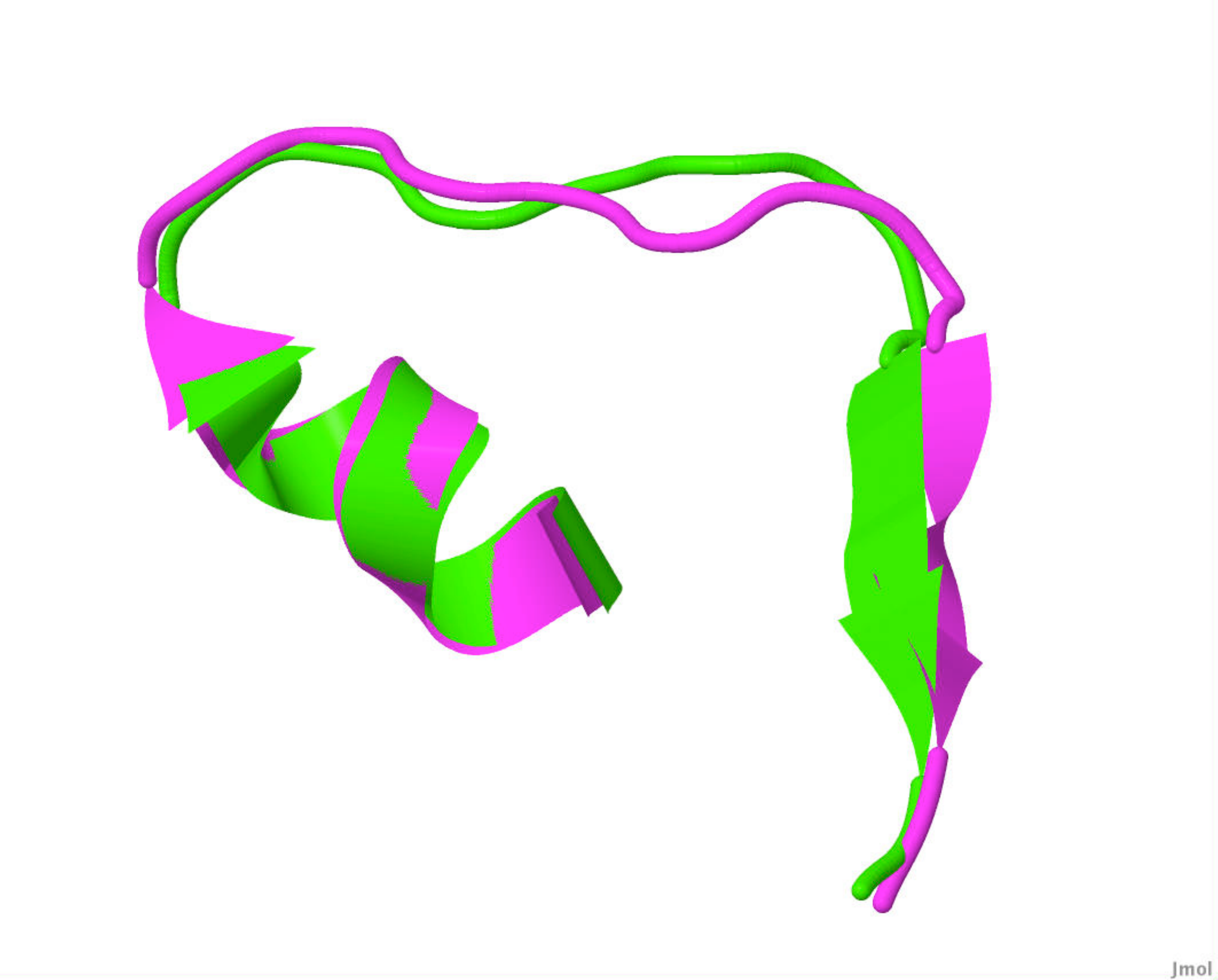}
\caption{Sites 398-416 in 3DLK (green; PDB indexing) and their approximation (purple) by (\ref{An2}), (\ref{ti2}) with parameter values given in Table II.
The RMSD distance is $\sim$ 1.13 \.A}\label{fig:02}
\end{figure}

 We conclude, that the present approach is suitable not only for long protein chains such as myoglobin, but it also describes 
 long loops and loops that connect very different kind of helices such as $\alpha$ helices, $3/10$ helices and $\beta$ strands.
 However, if the loop length increases substantially, we propose that a more accurate prescription is obtained by
 describing these loops as bound states of several short loops, each with  the profile (\ref{An2}), (\ref{ti2}). This is consistent
 with the well known fact that short supersecondary structures are known to recur many times in PDB proteins. A detailed analysis of
 long loops as bound states of short loops (multi-soliton states) will be presented elsewhere.

\section{Conclusion}

Using the myoglobin 1M6C as an example, we have demonstrated that the entire native fold of a long protein can be described 
with high accuracy as a combination of kink-solitons, in a manner that involves only one single elementary function. In this picture,
each of the solitons describe a loop configuration that interpolates between two different helices.   By inspecting a longer loop that
connects an $\alpha$-helix with a $\beta$-strand we have verified, that the approach remains valid with no loss of accuracy as the
loop size increases. However,
for substantially longer loops, we expect that an interpretation in terms of a multi-soliton configuration becomes more accurate both mathematically
and phenomenologically.
The parameters that characterize a particular protein fold are all global, and specific to its supersecondary helix-loop-helix motifs. 
Consequently the determination of these parameters becomes synonymous to a quantitative classification of proteins. The 
presence of an underlying Hamiltonian interpretation at the level of motifs also strongly suggests that our approach could eventually 
provide a bridge between comparative protein classification schemes such as CATH and SCOP,  and the  atomary level physics based 
approaches to protein folding and structure prediction, including folding  pathways and various other dynamical issues that presently 
can not  be easily addressed in qualitative protein classification schemes. This should open doors to new  ways of 
performing evolutionary, energetic and modelling studies.

\vskip 0.5cm 
\section*{Acknowledgement}
We thank D. van der Spoel and R. Lavery for discussions and comments.
This research has been supported by
the Vetenskapsr\.adet  grant  number 2009-4099.

%
%

\end{document}